\documentclass[structabstract]{aa}  
\usepackage{natbib}
\usepackage{graphicx}
\usepackage{txfonts}
\usepackage{color}

\begin{document}
\title{
Detecting scattered light from low-mass molecular cores at 3.6 $\mu$m
}
\subtitle{
Impact of global effects on the observation of coreshine
}

\author{
{J. Steinacker}\inst{1,2}
 \and
{M. Andersen}\inst{1}
 \and
{W.-F. Thi}\inst{1}
 \and
{A. Bacmann}\inst{1}
       }
\institute{
UJF-Grenoble 1 / CNRS-INSU, 
Institut de Planetologie et d'Astrophysique de Grenoble (IPAG) 
UMR 5274, Grenoble, F-38041, France\\
\email{stein@mpia.de}
\and
Max-Planck-Institut f\"ur Astronomie,
K\"onigstuhl 17, D-69117 Heidelberg, Germany
          }
\date{Received yesterday; accepted today}

  \abstract
{
Recently discovered scattered light at 3-5 $\mu$m from low-mass
cores (so-called "coreshine") reveals the presence of grains around 1 $\mu$m, which is larger than
the grains found in the low-density interstellar medium.
But only about half of the 100+ cores investigated so far show the effect.
This prompts further studies on the origin of this detection rate.
}
{   
We aim to supply criteria for detecting scattered light at 3.6
$\mu$m from molecular cloud cores.
}
{
From the 3D continuum radiative transfer equation, we derive the expected
scattered light intensity
from a core placed in an arbitrary direction seen from Earth. We use the
approximation of single scattering, consider extinction up to 2nd-order Taylor 
approximation, and neglect spatial gradients in the dust size
distribution.
We analyze how scattered light can outshine the absorbing effect of extinction 
in front of background radiation
by the core for given grain properties, anisotropic interstellar
radiation field and background field.
The impact of the directional characteristics of the scattering on the detection of scattered
light from cores is calculated for a given grain size distribution,
and local effects like additional radiation field components are
discussed. The surface brightness profiles of a core with a 1D density
profile are calculated for various Galactic locations, and the results are
compared to the approximate detection limits.
}
{
We find that for optically thin radiation and a constant size
distribution, a simple limit for detecting scattered light from a low-mass
core can be derived that holds for grains
with sizes smaller than 0.5 $\mu$m. The extinction by the core prohibits
detection in bright parts of the Galactic plane, especially near the
Galactic center. For scattered light received from low-mass cores with
grain sizes beyond 0.5 $\mu$m, the directional characteristics of the scattering
favors the detection of scattered light above and below the
Galactic center, and to some extent near the Galactic anti-center. We
identify the local incident radiation field as the major unknown causing
deviations from this simple scheme.
}
{
The detection of coreshine at a wavelength of 3.6 $\mu$m is
a complex interplay of the incident radiation, the extinction of
the background radiation, the grain properties, and the core properties like
sky position and mass.
}

\keywords{
 ISM: dust, extinction --
 ISM: clouds --
 Infrared: ISM --
 Scattering
         }

\authorrunning{Steinacker et al.}
\titlerunning{Detecting coreshine}
\maketitle

\section{Introduction}

Dense regions in molecular clouds have been identified as the site
where star formation starts. 
As the initial conditions impact the formation process, much effort
has been devoted to analyze the change in conditions
from the diffuse interstellar medium (ISM) to these dense cold cores
\citep[for an overview, see, e.g., ][]{2007ARA&A..45..339B}.
Through absorption, scattering and emission, cosmic dust grains embedded 
in the core gas enshroud the view on the star formation process
at optical wavelengths, but
also provide information on the physical conditions at longer wavelengths. 

A key quantity of the grains is their size distribution. 
For the diffuse ISM, \citet{1977ApJ...217..425M} were the first to suggest
a power-law distribution with a size limit of 0.25 $\mu$m for non-graphite grains
(hereinafter called "MRN(0.25)"). 
They emphasized that only weak
constraints could be derived for the larger grains in the distribution.
For the denser phases of the ISM, however, the modeling of thermal emission and extinction of grains
provided indirect evidence for the existence of a population
of grains beyond MRN(0.25)
\citep[e.g.][]{ 
2003A&A...398..551S, 2006MNRAS.373.1213K, 2006A&A...451..961R, 2008ApJ...684.1228S, 2009ApJ...690..496C, 2009ApJ...699.1866C, 2009ApJ...693L..81M}.
\citet{1996A&A...309..570L}
modeled the J, H, and K band scattered light images of the 
Thumbnail nebula and concluded that the limiting grain size exceeds 0.25 $\mu$m.

In a recent paper, evidence in the same direction
was reported for the core LDN183 by
\citet{2010A&A...511A...9S} based on the detection
of a weak extended surface brightness excess with respect to the large-scale
infrared background from the 
core visible in the 
3.6 and 4.5 $\mu$m IRAC bands of the {\it Spitzer} space telescope
called "coreshine".
Since the core is too cold for large grain thermal emission, and the
emission of stochastically heated particles like polycyclic aromatic hydrocarbons
or very small grains should be restricted to
the outer parts of the core to receive enough UV/optical energy, the 
radiation was interpreted as scattered interstellar light. 
Furthermore, despite the absence of strong stochastically heated emission in the
4.5 $\mu$m band, surface brightness excess was seen also in this band. The interpretation
as scattered light required a scattering efficiency of the dust that is only provided by
grains with radii up to 1 $\mu$m.

Although the central core of LDN183 is surrounded by slightly more mass than
most low-mass cores, its general conditions were not substantially different from that of
many other low-mass cores. Therefore, it was not entirely surprising that an investigation
of other cores in the IRAC bands led to the finding that about
half of the selected cores showed scattered light \citep{2010Sci...329.1622P}
\citep[see also][for a small subsample of cores with 3.6 $\mu$m excess]{2009ApJ...707..137S}.

A radiative transfer analysis of model cores revealed that the scattered light morphology
(the morphology referring to the gradients across the core, not the total surface brightness)
of low-mass cores should show no strong selection effect with position in 
the Galaxy aside from a weak enhancement towards the Galactic center (GC). For more
massive cores with 10 $M_\odot$, the morphology showed a crescent shaped asymmetry in the
morphology of the surface brightness excess.
A variation of the total expected flux as a function of sky
position was not investigated.

For a special region, namely the Gum/Vela region, \citet{2012A&A...541A.154P}
found a drastically reduced occurrence of scattered light (3 detections plus 3 uncertain
cases out of 24 objects). They did not attribute this absence to systematic
detection difficulties but to the action of a supernova remnant
blast wave which has affected the grain size population in this region.
The analysis of scattered light based on IRAC photometry
is compromised by PAH emission that in principle can contaminate all four IRAC
bands. This is true for any star forming region like $\rho$~Oph or Vela that is irradiated by close
medium or high-mass stars.
When scattered light and PAH emission regions
are spatially separated, one can use the almost PAH-free 4.5 $\mu$m band to distinguish
or use that the PAHs appear in all four IRAC channels opposite to the
scattered light being strongest in the 3.6 $\mu$m band and decreasing in flux with wavelengths.
When a region shows both scattered light and PAH emission, it is difficult to disentangle them.

The work presented in this paper focuses on deriving the general conditions
for the detection of scattered light at wavelength of 3.6 $\mu$m from a low-mass molecular cloud core.
Assuming various grain size limits, we explore the different
factors influencing the detection.
In Sect.~2 we derive approximative formula for the
coreshine detection limit from the radiative transfer equation. In Sect.~3 we compare
the formula with results from a grid of 3D radiative transfer calculations for
a model core spanning an all-sky map of coreshine surface brightness profiles. Observational constraints
are discussed in Sect.~4 for a set of example cores, and we conclude our findings in
Sect.~5.

\section{Deriving an approximate coreshine surface brightness formula}
\subsection{Optical depth and grain size distribution}
Determining the radiation observed from a molecular cloud core that is illuminated
by an external radiation field requires to run 3D radiative transfer calculations.
This is because almost all cores have a complex structure, because the external radiation field 
is strongly anisotropic, and the size distribution may show spatial variations. 
So far, only \citet{2010A&A...511A...9S} have modeled the
core LDN183 in such detail, but the large parameter space is difficult to probe
\citep[for a review of the computational challenge see][]{2013ARA&A..51...63S}.
In this section, we will discuss two approximations that reduce the
complexity of the detection problem while maintaining the essential physics.

\subsubsection{Optical depth approximation}
The first approximation is based on the fact that the surface brightness distribution
for the cores modeled so far scales with the column density measured, e.g., through
the 8 $\mu$m absorption profile. This is the case for optical thin scattered light.
Only a handful of the densest low-mass cores 
reveal profile depressions in the central core region. 
Model calculations for the low-mass
cores LDN183 \citep[][ with depression]{2010A&A...511A...9S}
and LDN260 \citep[][ without depression]{2013A&A...559A..60A} indicate that the depression is
caused by optical depth effects. For the majority of cores, the assumption of optically thin
scattering and absorption is therefore a reasonable approach.
In Sect.~2.2 we will discuss the limits of this assumptions.

The general stationary continuum radiative transfer equation has the
form of an integro-differential equation where the scattering integral contains
the unknown intensity itself and advanced solution methods like Monte-Carlo or
ray-tracing are required \citep[see][]{2013ARA&A..51...63S}. 
Since the model calculations performed for \citet{2010Sci...329.1622P} indicate that 
almost all cores scatter optically thin at 3.6 $\mu$m, we 
simplify the equation by assuming single scattering.
We note that the contribution of the scattering to the extinction depends on
the grain size distribution. This means that the limits for using single scattering
and approximating the exponential extinction term by terms from the Taylor expansion
varies with the grains sizes as discussed in Sect.2.2.

The
3.6 $\mu$m surface brightness seen from model cores placed at a constant distance
around the observer is
\begin{eqnarray}
I(y,z)
&=&
e^{-\tau_{fg}(y,z)}
\left[ 
 I_{bg}(y,z)\ e^{-\tau_{ext}(y.z)}+I_{fg}(y,z)+
\right.
\cr
& &
\left. 
 +\int\limits_{-\infty}^{+\infty}dx
  \int\limits_{a_1}^{a_2}da\ n(a,x,y,z)\ \sigma_{sca}(a,x,y,z)\ \left< I_{in}\right>_\Omega
\right]
\end{eqnarray}
with the direction-averaged incident radiation field 
\begin{eqnarray}
\left<I_{in}\right>_\Omega&&=\cr
&&
\int\limits_{\bigcirc}d\Omega\ p(\Omega,a,x,y,z)\ I_{ISRF}(\Omega)\ 
e^{-\tau_{ext;1}(\Omega,x,y,z)}\ 
e^{-\tau_{ext;2}(x,y,z)}.
\end{eqnarray}
$I_{bg}$ denotes the background surface brightness, 
$\tau_{ext}$ is the optical depth due to extinction through the core along the line-of-sight (LoS), 
and $a_1$ and $a_2$ are the lower and upper limit for the grain size distribution
$n$.
Furthermore, $\sigma_{sca}$ stands for the scattering cross section,
$I_{fg}$ is the foreground surface brightness, 
$\tau_{fg}$ is the optical depth for foreground extinction, 
and
$p$ is the normalized phase function of the grains for scattering from any direction towards the observer.

\begin{figure}
\includegraphics[width=9cm]{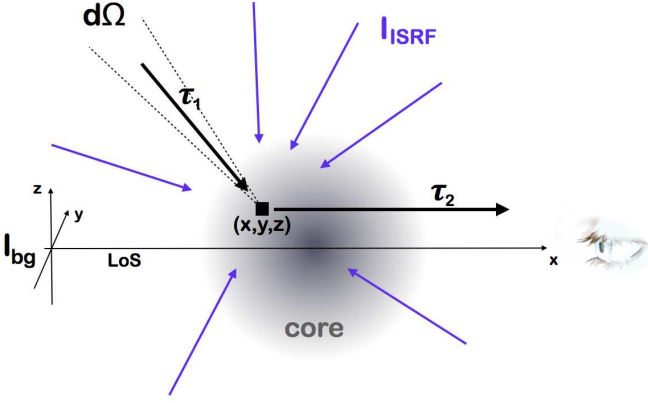}
\caption{
Illustration of the choice of coordinates following the scattering of
radiation in a core. $\tau_{ext;1}$ and $\tau_{ext;2}$ are the optical depth
for extinction for incident radiation and radiation being scattered towards the
observer, respectively.
        }
\label{sketch}
\end{figure}

Interstellar radiation on its way to the core position $(x,y,z)$ 
from a certain solid angle $\Omega$ undergoes extinction
described by $\tau_{ext;1}$.
The radiation that is scattered by the dust at $(x,y,z)$ and leaves towards the observer
again undergoes extinction with the optical depth $\tau_{ext;2}$.
The Cartesian directions are $x$ pointing from the core to the observer,
$y$ along the Galactic coordinate $l$ in the Plane of Sky (PoSky), and
$z$ perpendicular to $y$ in the PoSky. The set-up is illustrated in Fig.~\ref{sketch}.

\subsubsection{Approximation of the grain size distribution}
The second approximation is also motivated by the smooth brightness distribution 
of most low-mass cores with coreshine. The images derived for LDN183
in \citet{2010A&A...511A...9S}
used spatial gradients in the grain size distribution
and the resulting surface brightness gradients were steeper than in the observed
images. 
We therefore assume here like in \citet{2013A&A...559A..60A} for the core
LDN260 and in \citet{L1506C} for LDN1506C a spatially constant size distribution
\begin{equation}
n(a,x,y,z)=s(a)\ t(x,y,z).
\end{equation}
with the size distribution $s$ and the spatial distribution $t$.\footnote{
We note that this not necessarily excludes locally acting grain growth mechanisms
proposed for cores
\citep{1993A&A...280..617O,1994ApJ...430..713W,2011A&A...532A..43O},
as long as the turbulence that creates the relative motions of the grains also
provides turbulent mixing to smooth out
spatial gradients in $n$.} 
The optical depth for extinction then takes the form
\begin{eqnarray}
\tau_{ext}(y,z)&=&
\int\limits_{-\infty}^{+\infty}dx\ 
\int\limits_{a_1}^{a_2}da\ s(a)\ t(x,y,z)\ \sigma_{ext}(a)\cr
&=&
\left<\sigma_{ext}\right>_s\ N(y,z)
\end{eqnarray}
with the column density 
\begin{equation}
N(y,z)=\int\limits_{-\infty}^{+\infty}dx\ 
t(x,y,z)
\end{equation}
and the size distribution-averaged extinction cross section
\begin{equation}
\left<\sigma_{ext}\right>_s=
\int\limits_{a_1}^{a_2}da\ s(a)\ \sigma_{ext}(a).
\label{sext}
\end{equation}
\subsection{Coreshine surface brightness using first-order Taylor terms}
In the following, we will neglect the optically thin
foreground extinction $\exp(-\tau_{fg})$
that attenuates the flux of the radiation finally leaving the core towards the observer.
This additional extinction may be caused
by a core surrounding cloud or by diffuse dust not related to the core hosting cloud. 
The additional extinction decreases the detection rate of coreshine for distant
or deeply embedded low-mass cores since it moves the ratio of coreshine intensity and
background intensity towards the detection limit of the instrument.

We expand the extinction factor $\exp(-\tau)\approx 1-\tau$ as Taylor series in first-order 
around $\tau=0$.
Considering single scattering, a constant grain size distribution, 
and only expressions up to the order $\tau$
the surface brightness for radiation $I$ along the line-of-sight is
\begin{eqnarray}
I(y,z) &-& I_{bg}(y,z) - I_{fg}(y,z) \approx\cr
&-&\left<\sigma_{ext}\right>_s\ N(y,z)\ I_{bg}(y,z)\cr
&+&
N(y,z)\ 
\int\limits_{a_1}^{a_2}da\ s(a)\ 
\sigma_{sca}(a)\ \int\limits_{\bigcirc}d\Omega\ p(\Omega,a)\ I_{ISRF}(\Omega).
\label{avI}
\end{eqnarray}
The term that prevents
combining the integrals over the grain sizes and the solid angles
is the phase function $p(\Omega, a)$ depending on both quantities.

\begin{figure}
\includegraphics[width=9cm]{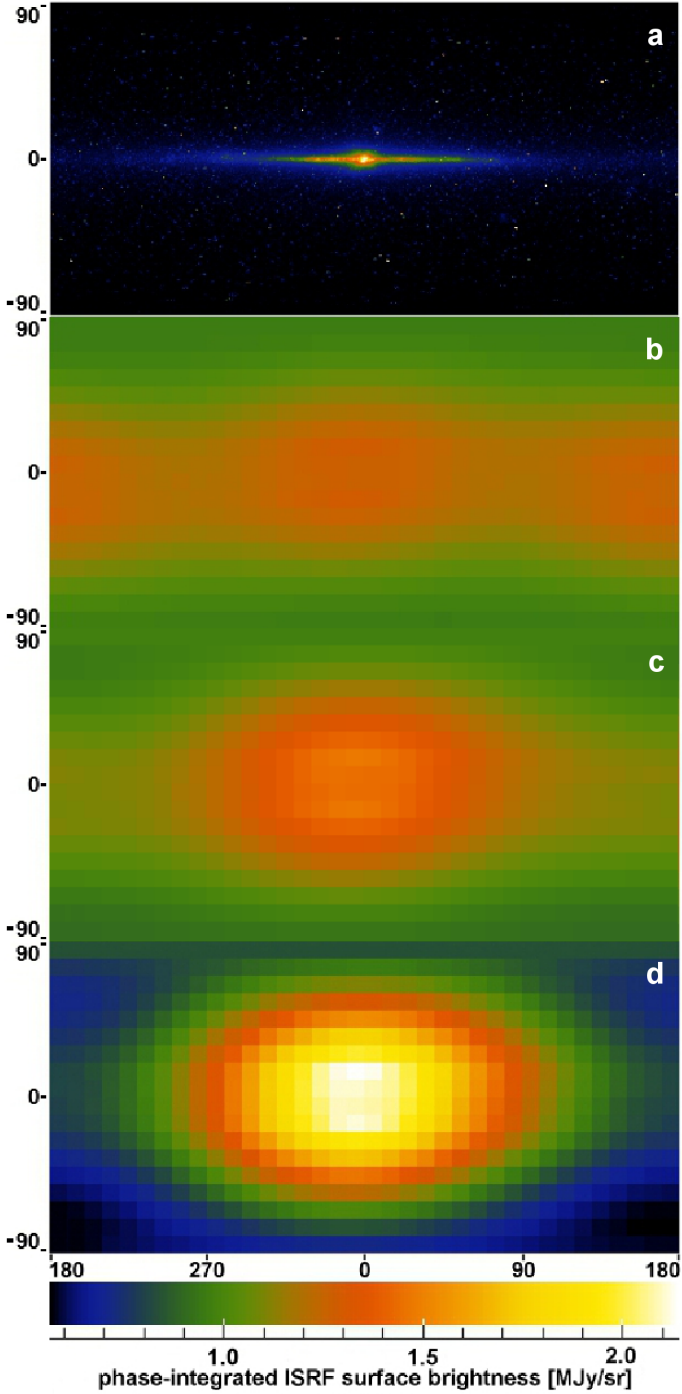}
\caption{
a) DIRBE allsky map used as incident radiation field with surface brightnesses
ranging from 0.08 to 12.6 MJy/sr.
b-d) Three all-sky maps in Galactic coordinates centered on the GC.
The maps show the phase function-integrated surface brightness of ISRF radiation being
completely scattered by a cell with dust at each sky location towards the observer.
The maximal grain size of the dust size distribution is 0.25 (a), 0.5 (b), and 1 $\mu$m (c),
respectively.
Local maxima arise near the Galactic plane (especially near the Galactic
Center and the anti-center) due to peaks in the phase function for forward
and backward scattering.
        }
\label{p3}
\end{figure}
\begin{figure}
\includegraphics[width=9cm]{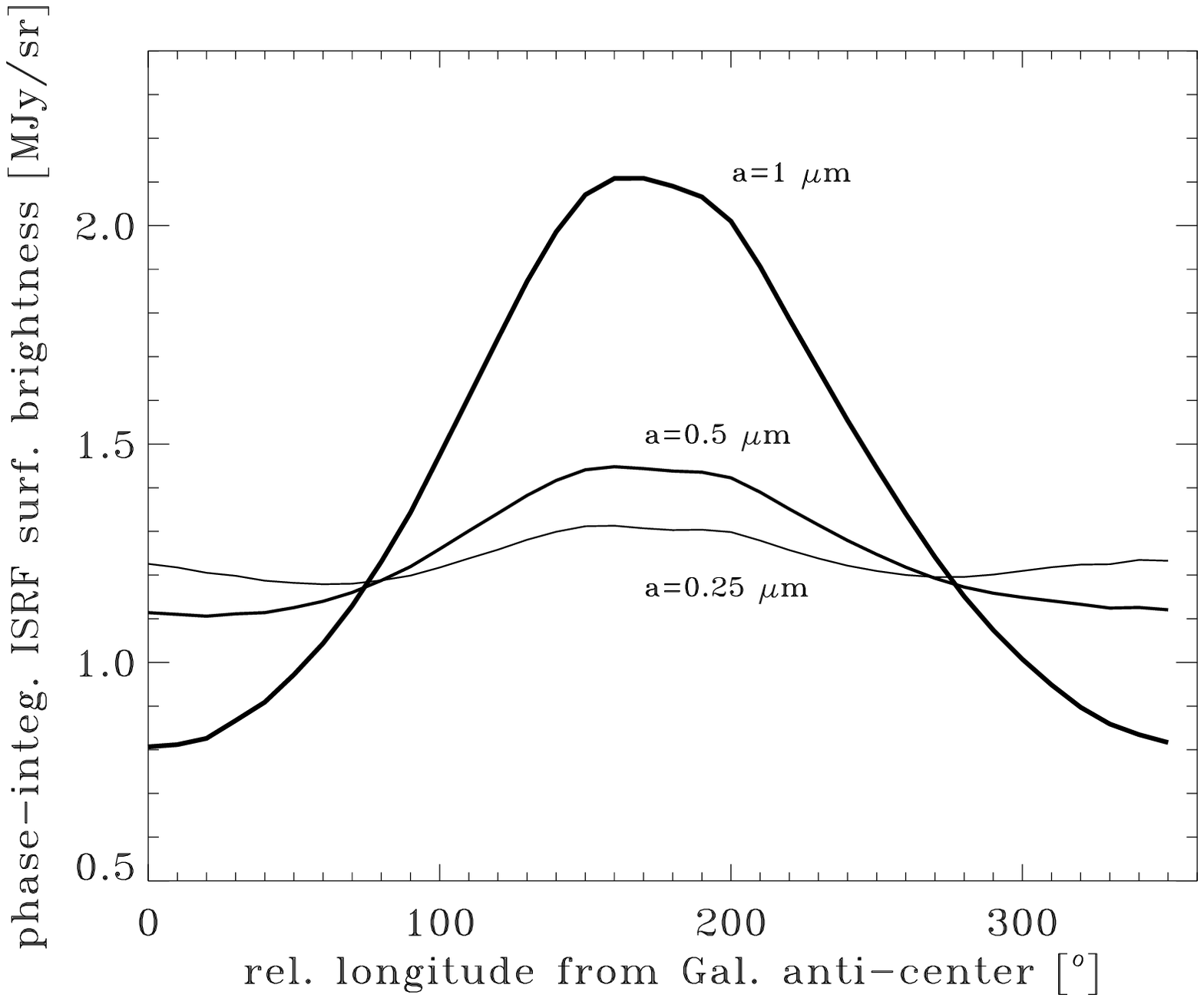}
\caption{
Horizontal cut through the phase-integrated surface brightness 
shown in Fig.~\ref{p3} in the Galactic midplane for the three 
maximal grain sizes 0.25, 0.5, 1 $\mu$m.
        }
\label{pp}
\end{figure}
The overall effect of the phase function is to narrow the scattering angle distribution
once $2\pi a > \lambda$. The strong beaming of the phase function is shown in polar coordinates
for spherical homogenous silicate grains of various sizes at $\lambda =$ 3.6 $\mu$m 
in Fig.~7 in \citet{2010A&A...511A...9S}.
It is evident that the phase function is not constant even for small grains.

To explore the impact of the phase function on the coreshine, we assume a size 
distribution, an ISRF, and dust opacities.
We consider a grain size distribution MRN($a_2$).
The impacting radiation field $I_{ISRF}$ is approximated
by the zodiacal light-subtracted 3.5 $\mu$m Diffuse Infrared Background Experiment (DIRBE) map provided by
LAMBDA\footnote{http://lambda.gsfc.nasa.gov}.
A local component of the ISRF in addition to the ISRF is discussed below.
The grain opacities are taken from
\citet{1984ApJ...285...89D}.
The phase function is usually given as a probability for scattering from
one direction into another as a function of the scattering angle between the two directions
\citep[see, e.g., as used in this work,][]{2003ApJ...598.1017D}.
But rather than integrating over the scattered
radiation in all directions, we are here interested in integrating the radiation that is scattered
into one direction (the LoS) and coming from all directions.
The simplest choice is a spherical grid where the solid angles (the weights of the integration)
are calculated from the spherical $\Delta\Omega$ attached to the direction in discretized spherical
coordinates. A disadvantage is that this grid increases resolution at the poles and that might not 
always be the direction where the integrand has its strongest gradients.
Therefore, also direction distributions with equal solid angles are in use.
\citet{1996JQSRT..56...97S} have proposed to calculate nearly isotropically distributed direction grids and
solid angles with a Metropolis optimizer for any number of grid points.
Alternatively, a HEALpix distribution can be used which uses constant solid angles for specific numbers
of grid points
\footnote{http://healpix.jpl.nasa.gov/} (used, e.g. for the DIRBE data).
Monte-Carlo radiative transfer solvers, in turn, do not discretize the direction space and choose
a random direction from a probability function determined by the integrand.
Since our main radiative transfer code uses a spherical grid, we have kept this choice for practical reasons.

Fig.~\ref{p3} shows the allsky map used as incident radiation and three maps of
\begin{equation}
P(\Omega_{cell},a)=
\int\limits_{\bigcirc}d\Omega\ p(\Omega,a)\ I_{ISRF}(\Omega)
\end{equation}
for a dust cell placed at any direction $\Omega_{cell}$
in the sky for the grain sizes $a_2$=0.25, 0.5, and 1 $\mu$m (bottom).
The strongly anisotropic ISRF of the Galaxy is smoothed out with remaining gradients
increasing with maximal grain size.
The gradients can also be inferred from the profiles along horizontal cuts through the images
in the midplane as shown in Fig.~\ref{pp} for the three maximal grain sizes.

For grains up to 0.5 $\mu$m the variations are of the order of 30\% and
we therefore first discuss the detection of coreshine for grains up this size
limit using the approximation 
$<p I_{ISRF}>_\Omega \approx <I_{ISRF}>$, and
will later explore the usage of phase functions with stronger beaming for
grains with sizes beyond 0.5 $\mu$m.
Eq.~(\ref{avI}) then for each location (y,z) in the PoSky has the form
\begin{equation}
I - I_{bg} - I_{fg}
\approx
N\ 
\left(
\left<\sigma_{sca}\right>_s\ \left<I_{ISRF}\right>_\Omega-\left<\sigma_{ext}\right>_s\ I_{bg}
\right)
\label{simple}
\end{equation}
\citep[compare to][Eq. (8), their formula adopts the background, but they neglect 
it for the Thumbprint nebula]{1996A&A...309..570L}.

Without incident radiation, the core is seen in extinction against the background 
radiation with the intensity $I_{bg}$ being absorbed and scattered out of
the LoS. When increasing $I_{ISRF}$, the effect of extinction of $I_{bg}$ is in part
compensated by radiation scattered into the LoS. The profile of the surface brightness
flattens until enough
scattered light along the LoS is created to compensate for the extinction.

\begin{figure}
\includegraphics[width=9cm]{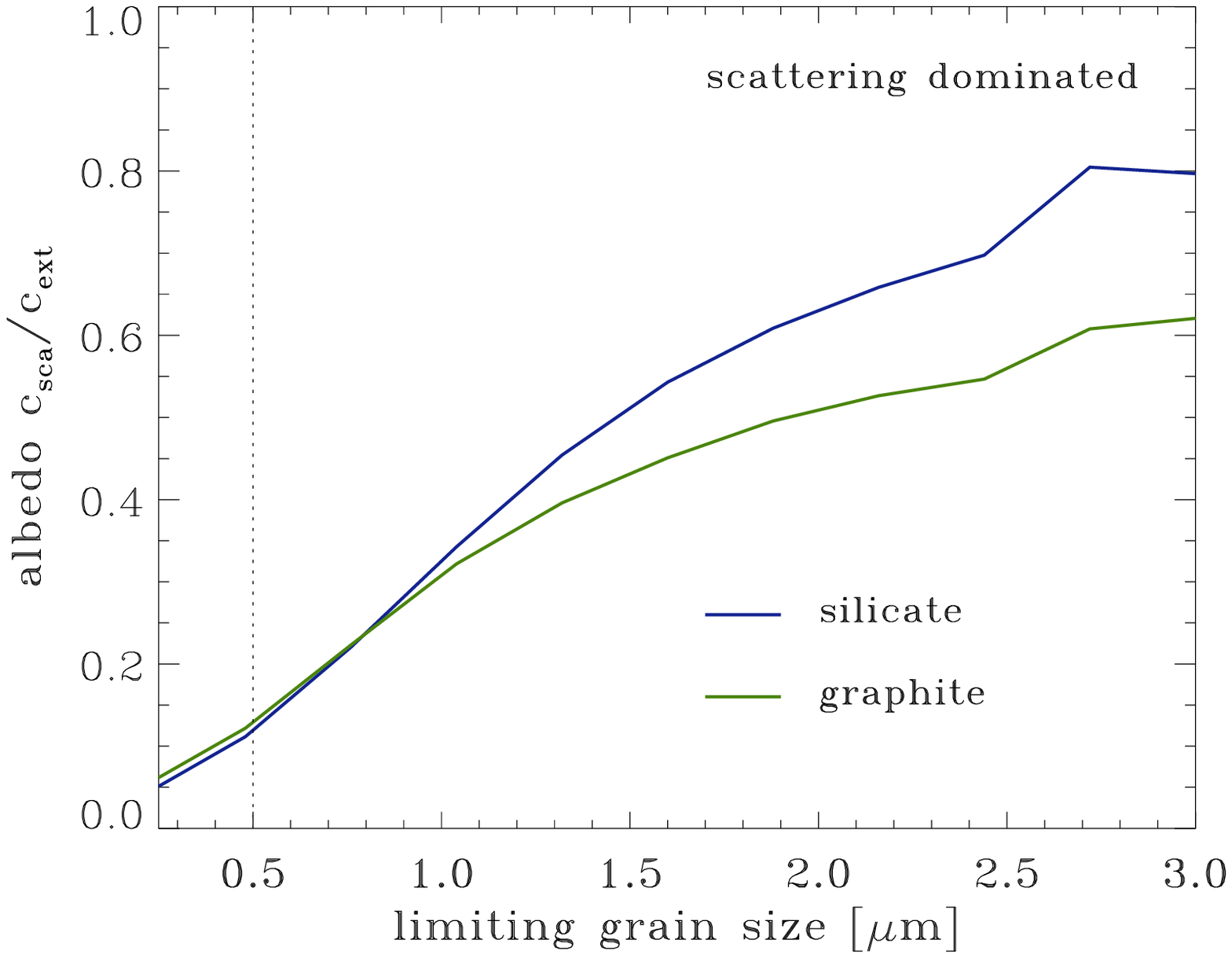}
\caption{
Albedo based on the size-averaged 
scattering and extinction cross section assuming an MRN distribution 
as a function of the maximum grain size 
for silicate and graphite \citep[][]{1984ApJ...285...89D}. The
dotted line indicates the limiting grain size 
where the error due to neglecting the phase function in the 
scattering integral is less than 30\%.
        }
\label{sigma}
\end{figure}

In the following, we call this case 
"scattering-extinction balance".
Its condition
\begin{equation}
\frac{I_{bg}} {\left<I_{ISRF}\right>_\Omega}
=
w
=
\frac{\left<\sigma_{sca}\right>_s}{\left<\sigma_{ext}\right>_s}
\label{con}
\end{equation}
on the right hand side only depends on the properties of the grain size distribution
described by the albedo 
$w$.
In Fig.~\ref{sigma}, the albedo is plotted as a function of the upper
limit for the grain radius for the MRN size distribution and the silicate and
graphite opacities.
The total dust mass is the same for all curves.
The
dotted line indicates the chosen grain size limit for
assuming $P\approx <I_{ISRF}>$.

Since the albedo is always smaller than 1, a 
scattering-extinction balance can only occur
when 
\begin{equation}
I_{bg}\ <\ \left<I_{ISRF}\right>_\Omega.
\label{flat}
\end{equation}

Moreover, the smaller the grain size limit, the more incident radiation
is needed to reach at least the scattering-extinction balance condition or observe coreshine.
In the case of a typical diffuse ISM MRN(0.25), the incident field can create coreshine above the
balance when it is more than about 20 times stronger than the background
field.

This implies the interesting conclusion that coreshine can be observed
even for MRN(0.25). 
But the enhancement of the ISRF of about 20 over the background
is rarely met if no stellar source is located in the immediate vicinity 
of the core, or the core is part of a massive star formation region.

A complication is a background variation across the core. 
This can not be compensated by subtracting
a function describing the background. The reason is that the background also
enters the scattering-extinction balance condition causing regions of suppressed flux
in the coreshine surface brightness morphology of the core. For the following simple
considerations we will neglect background gradients across the core.

In order to investigate the regions in the sky for which coreshine would be observable
according to the scattering-extinction balance condition, we have to determine $I_{bg}$, 
$\left<I_{ISRF}\right>_\Omega$, and define the grain
size distribution.\\ 
(a) $\left<I_{ISRF}\right>_\Omega$: 
We assume that the radiation field impacting the cores can be approximated
by the DIRBE 3.5 $\mu$m once the contamination from stars is removed. 
This is only to be modified when local effects like absorption by a nearby 
core or filament segment or illumination by neighboring stars or photodissociation
regions alter the local radiation field.\\
(b) s(a): We use an MRN distribution with a sharp cut-off at a maximal grains radius $a_2$. 
The phase function starts to impact the scattering integral with deviations 
beyond 30\% \citep[assuming][opacities]{1984ApJ...285...89D} for grains larger than 0.5 $\mu$m
and needs to be considered in the scattering-extinction balance condition for distributions with grains beyond this size.\\
(c) $I_{bg}$: There is no absolutely calibrated map of the 3.6 $\mu$m sky with the resolution of a 
few beams
across a low-mass core of typical diameter 150$\arcsec$ with distances $< 500$ pc. The WISE all-sky maps only 
provide relative photometric measurements. Akari only provides 
relative photometric measurements all-sky maps at 9 $\mu$m or larger.
The all-sky 3.5 $\mu$m DIRBE map remains the highest-resolving map 
to be used. However, this map might not be a realistic representation of
the background field behind the core for two reasons.
First, the DIRBE map consists of large Healpix 
segments\footnote{http://healpix.jpl.nasa.gov/} with an average size of 
0.35$^\circ$ $\times$ 0.23$^\circ$ which is smaller than the observing beam
where all background variations behind the cores with
typical scales of about 150\arcsec\ are averaged to a single value.
Second, the DIRBE map contains the stellar component which does not enter the coreshine
modeling. This overestimates the background, and the stellar component must be
subtracted using the WISE data.

\begin{figure*}
\includegraphics[width=18.5cm]{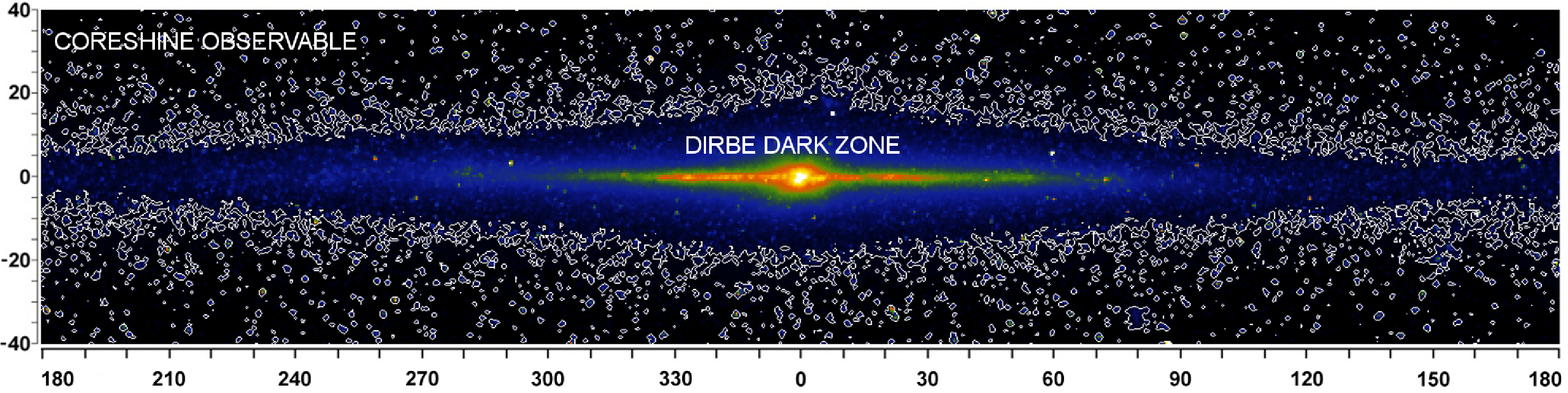}
\includegraphics[width=18.5cm]{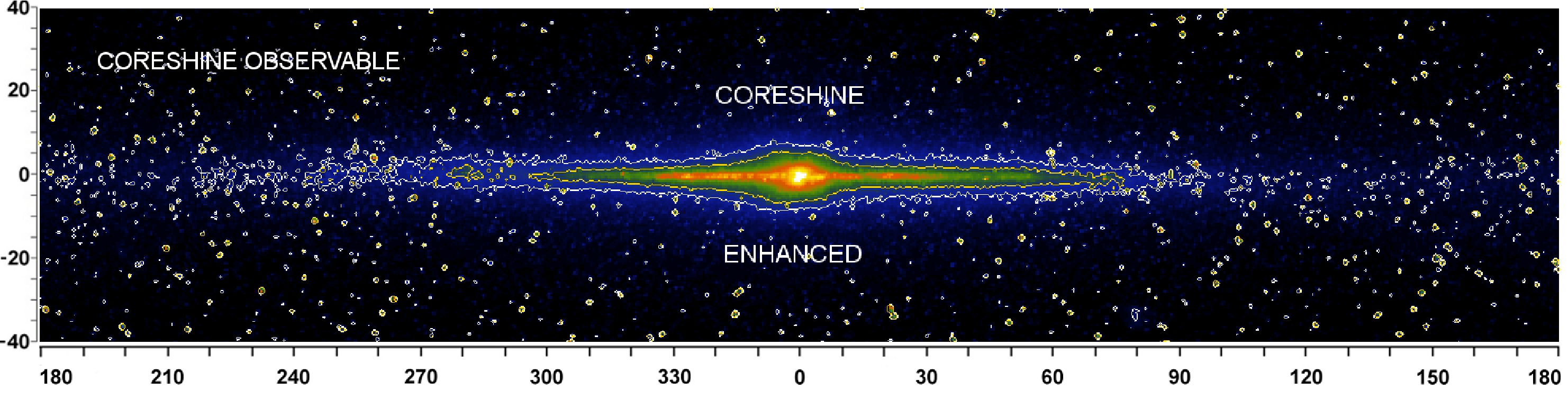}
\includegraphics[width=18.5cm]{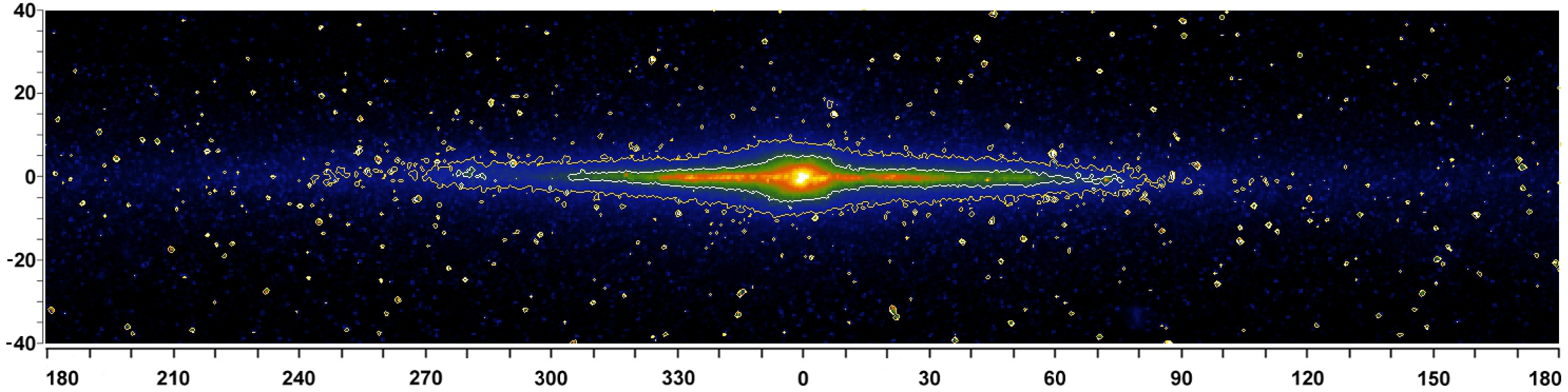}
\caption{
Color-coded is the DIRBE all-sky map between latitudes of -40$^o$ to +40$^o$ 
at 3.5 $\mu$m which serves as background after subtracting the stellar contamination.
Overlayed are contours of the scattering-extinction balance for MRN(0.5) (top), 
for MRN(1) taking into account phase function effects and treating extinction
in 1st order Taylor approximation (middle), and for MRN(1) and 2nd-order Taylor
approximation (bottom), respectively.
        }
\label{Flatline}
\end{figure*}
Fig.~\ref{Flatline} (top panel) shows the scattering-extinction balance distribution for 0.5 $\mu$m grains
as white contours on top
of the DIRBE all-sky map (-40$^\circ$ to +40$^\circ$ in latitude). 
The map shows two general features. 
There is an inner part without contour lines. In this "DIRBE dark zone" the background
exceeds the mean interstellar radiation field and no coreshine is expected
according to Eq.~(\ref{flat}).
The zone encloses most of the Galactic plane and especially the GC. 
Outside the zone the contour map shows substantial variation on small scales so that 
local effects dominate whether coreshine is observable. In the outer parts of the map
the overall decreasing background makes coreshine detection more likely again.
The DIRBE map overestimates the background due to the stellar content. Hence, we have
approximately corrected by dividing the DIRBE flux by a number that was derived as follows.
We investigated the background given by DIRBE for a few cores subtracting the
stellar content based on the WISE maps \citep{2013A&A...559A..60A}, and found that the fluxes are reduced 
by a factor of the order of 2. This is the
value we used here.
We note that a correct subtraction of the stellar component would reduce the gradients
in the background and therefore in the shown contours for the scattering-extinction balance
distribution but not modify the global separation between dark zone, variation region, and 
range with possible detection of coreshine.

Both the modeling with spatial grain size variations by
\citet{2010A&A...511A...9S}
and with spatially constant grain size distributions by
\citet{2013A&A...559A..60A}
found that the
modeling of the core surface brightnesses require the presence of larger grains.
Therefore, we relax our assumptions and allow for grains larger than 0.5 $\mu$m, which
requires allowing for a directional dependence of the
phase function.
The scattering-extinction balance condition Eq.~(\ref{con}) then becomes
\begin{equation}
\left<\sigma_{sca}\ \left<p\ I_{ISRF}\right>_\Omega\right>_s
=
\left<\sigma_{ext}\right>_s\ I_{bg}
\label{con1}.
\end{equation}
In this case, the location of the core enters both by the local background
and the global relation of core position and anisotropic illumination by the
ISRF via the phase function. Moreover, the increased grain size moves
the albedo closer to unity as visible in Fig.~\ref{sigma}.

\begin{figure}
\includegraphics[width=9.25cm]{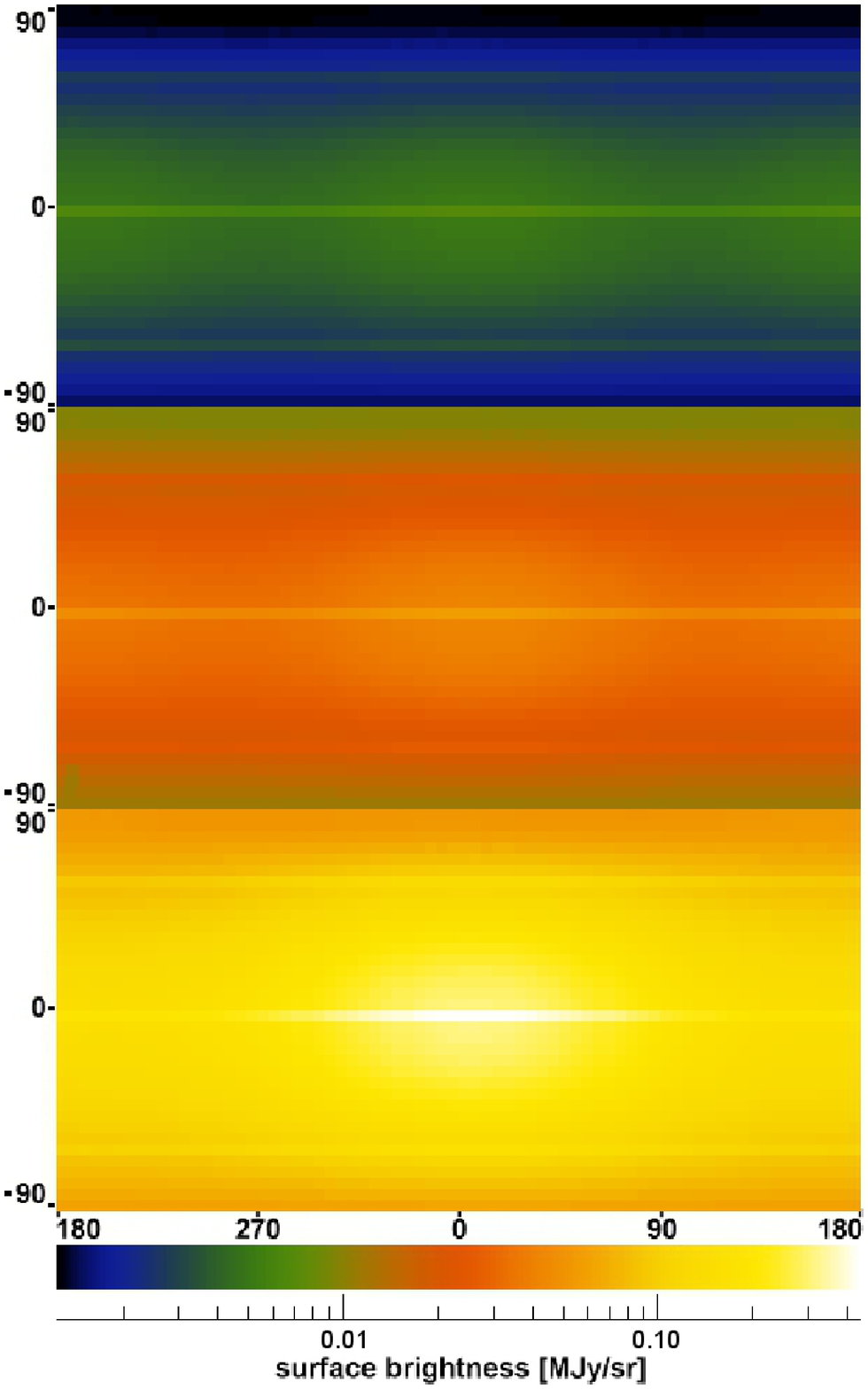}
\caption{
The impact of the phase function as a global effect on the observation of
coreshine:
Shown are three all-sky maps of the scattered light surface brightness from a 
1 $M_\odot$ core
that is placed in all directions seen from Earth. 
Three limiting grain sizes are considered (top
to bottom): 0.25, 0.5 and 1 $\mu$m.
        }
\label{integ3}
\end{figure}
The effect of allowing larger grains size limits of the grain distribution is two-fold:

(a) The phase function directional dependence.
The scattering integral in Eq.~(\ref{avI}) folds the phase function with its two maxima
at forward and backward scattering with the band-like ISRF direction morphology of the
DIRBE map and boosts the contribution weighted with the scattering efficiency. 
The resulting scattering integral contribution from a standard 1 $M_\odot$ core
 is shown in Fig.~\ref{integ3} as three all-sky maps in
Galactic coordinates
centered on the GC for the limiting grains sizes 0.25, 0.5 and 1 $\mu$m (top to bottom).
The amplification due to the phase function and the scattering cross section
make the scattering integral largest at the GC and its vicinity.
(b) The enhanced scattering due to big grains.
The larger grains increase the albedo shown in Fig.~\ref{sigma}
thus allowing for larger background values to still meet the scattering-extinction balance
condition for a given incident radiation field. This shrinks the DIRBE darkzone and coreshine
can be observed in a larger portion of the sky compared to the coreshine detection for
smaller grains.
The corresponding balance contours are shown in Fig.~\ref{Flatline} (middle)
on top of the DIRBE map (white contours for the DIRBE map used as background, yellow
for using a DIRBE map with half the intensity that corrects for the stellar contribution).

\begin{figure}
\includegraphics[width=9cm]{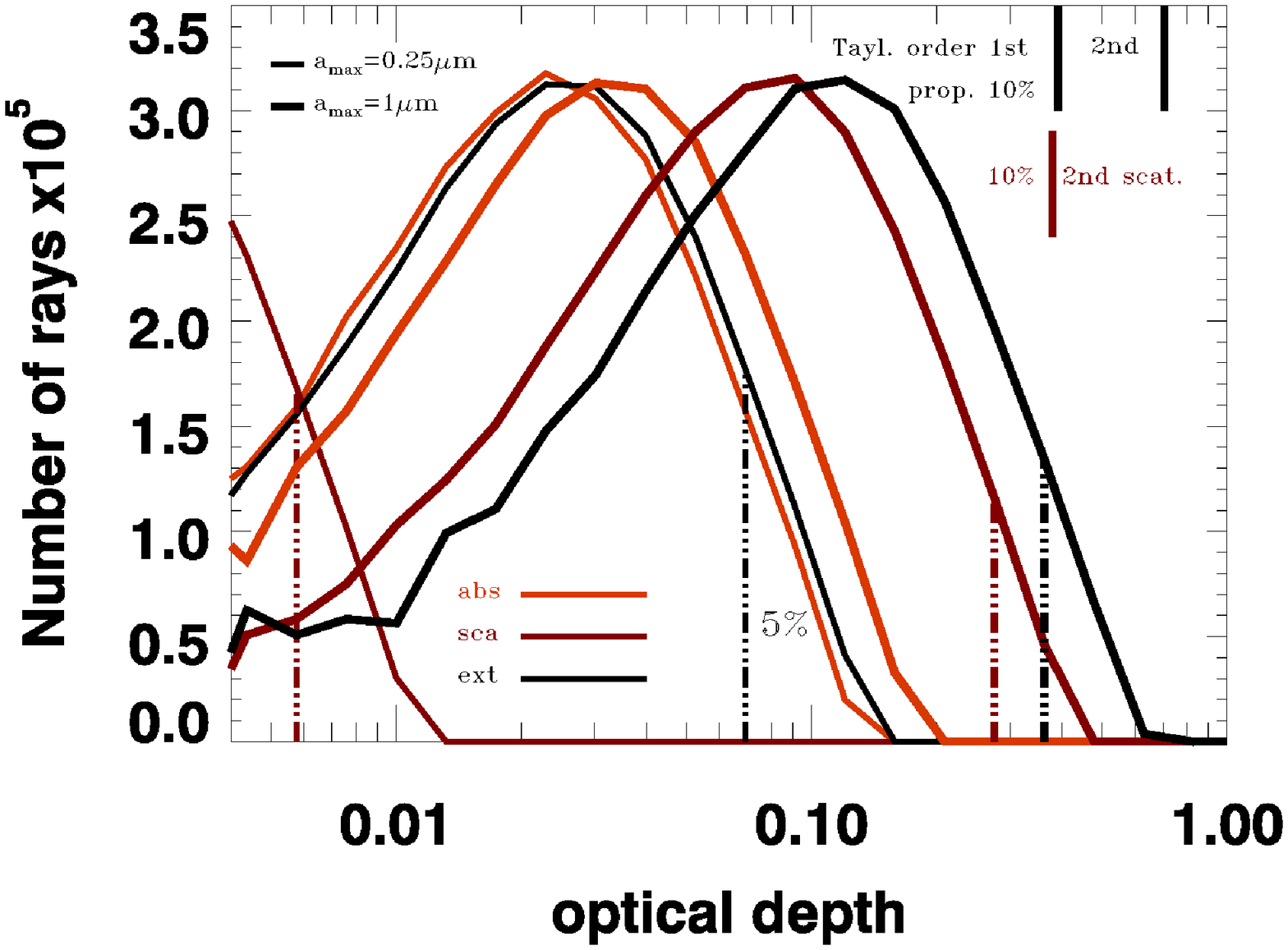}
\caption{
Number distribution of rays through a standard core as a function of
the optical depth for absorption (red), scattering (pink),
and extinction (black), respectively.
The thin and thick lines corresponds to MRN(0.25) and MRN(1) distributions, respectively.
The dash-dotted line give the limiting $\tau$ for which 5\% of all rays have
larger $\tau$.
The three vertical bars in the upper left indicate the $\tau$ values
for which the relative error in the Taylor expansion of the 
extinction in first and second order reaches 10\%, and the $\tau$
where the probability for second scattering reaches 10\%.
}
\label{Tayl}
\end{figure}

To explore the limits of the validity of the 
underlying approximations of optically thin scattering 
$\tau_{sca}\ll 1$ and the 1st- and 2nd-order Taylor expansion of $\exp{-\tau_{ext}}$, 
we have calculated 
the optical depth for absorption and scattering for about 3 million rays through a standard core
with a mass of 1 $M_\odot$ with about 600 direction per spatial point.
In Fig.~\ref{Tayl} we show the number distribution of rays per optical depth for
absorption (red), scattering (pink), and extinction (black), respectively.
The thin lines represent a core with an MRN(0.25) size distribution. Since all grains
are in the Rayleigh limit at this wavelength, scattering predominantly occurs 
with smaller optical depths than absorption and 
the absorption and extinction distribution are close to each other.
We also indicate the limit in $\tau_{ext}$ where only 5\% of all rays have a larger
optical depths
as vertical dash-dotted line.
For an MRN(1) distribution, scattering dominates over absorption, and the
distribution is moved towards larger optical depth values.
In the right upper corner we indicate three limiting $\tau$-values by vertical bars:
at about $\tau$=0.4, the relative 1st-order Taylor expansion error 
$[|\exp{-\tau_{ext}}-(1-\tau)|]/\exp{-\tau_{ext}}$ reaches 10\%,
at about $\tau$=0.7, the relative 2nd-order error reaches 10\%,
and slightly below $\tau$=0.4 
the probability for second scattering 
$(1-\exp{-\tau_{sca}})^2$ reaches 10\%.
Compared with the ray number distributions for MRN(1), about 5\% of all rays 
encounter extinction where the probability for second scattering is above 10\%
and where the first-order approximation of the extinction fails by 10\%.
Due to the centrally-peaked profile of the core, these are rays through the central
part of the core. For cores larger than the assumed 1 $M_\odot$, this will be the
region where deviations from the derived extinction-scattering balance
will appear first.

\subsection{Coreshine surface brightness using second-order Taylor terms}
For optical depths beyond $\tau=$0.4 (see Fig.~\ref{Tayl}) caused by 
larger column densities,
second-order Taylor terms for the approximation of $\tau$ appear in the 
balance condition.
Neglecting third-order terms, the surface brightness for radiation along the line-of-sight becomes
\begin{eqnarray}
&&I(y,z)- I_{bg}(y,z) - I_{fg}(y,z) \approx
-\left<\sigma_{ext}\right>_s\ N(y,z)\ I_{bg}(y,z)\cr
&&+N(y,z)\left<\sigma_{sca}\left<p\ I_{ISRF}\right>_\Omega\right>_s
+\frac{1}{2}\left<\sigma_{ext}\right>_s^2\ N(y,z)^2 I_{bg}(y,z)\cr
&&-\left<\sigma_{sca}\ \sigma_{ext}\ \left<p\ I_{ISRF}\ D\right>_\Omega\right>_s
\end{eqnarray}
with the directional transmission function
\begin{equation}
D=
\int\limits_{-\infty}^{+\infty}dx\ 
\left[ N_1(\Omega,x,y,z)+N_2(x,y,z) \right] t(x,y,z)
\label{2ndTayl}
\end{equation}
depending only on the spatial density structure, 
the column density $N_1$ from a point $(x,y,z)$ in the direction $\Omega$ and
$N_2$ from the point to the observer.
Two new terms arise compared to Eq.~(\ref{avI}). 
The first term $$\frac{1}{2}\left<\sigma_{ext}\right>_s^2\ N(y,z)^2\ I_{bg}(y,z)$$
is a positive $\tau^2$-correction to the extincted background radiation, and
reduces the background impact on the observed profile.
The second term corrects for extinction effects of the incident radiation 
field.

The scattering-extinction balance condition in second-order approximation for grains with $a<0.5$ $\mu$m
now reads
\begin{eqnarray}
&&\frac{\left<\sigma_{ext}\right>_s}{\left<\sigma_{sca}\right>_s}
\left( 1 - \frac{1}{2}\left<\sigma_{ext}\right>_s N \right) =\cr
&&\frac{\left<I_{ISRF}\right>_\Omega} {I_{bg}} -
\frac{
\left<\sigma_{sca}\ \sigma_{ext}\ \left<I_{ISRF}\ D \right>_\Omega \right>_s}
{\left<\sigma_{sca}\right>_s\ I_{bg} N}.
\label{FL2}
\end{eqnarray}
For small $\tau_{ext}=\left<\sigma_{ext}\right>_s N$, Eq.~(\ref{FL2}) is similar to the 1st-order
condition Eq.~(\ref{con}) on the left hand side of the equation.
The impact of the directional coupling via the extinction of the incident radiation 
described by the directional transmission function, however, strongly depends
on the core structure and its position in the sky with respect to the GC.
A density enhancement towards the GC, e.g., can reduce the incident
radiation so that the correction term containing $D$ impacts the balance condition by shielding.

\begin{figure}
\includegraphics[width=9cm]{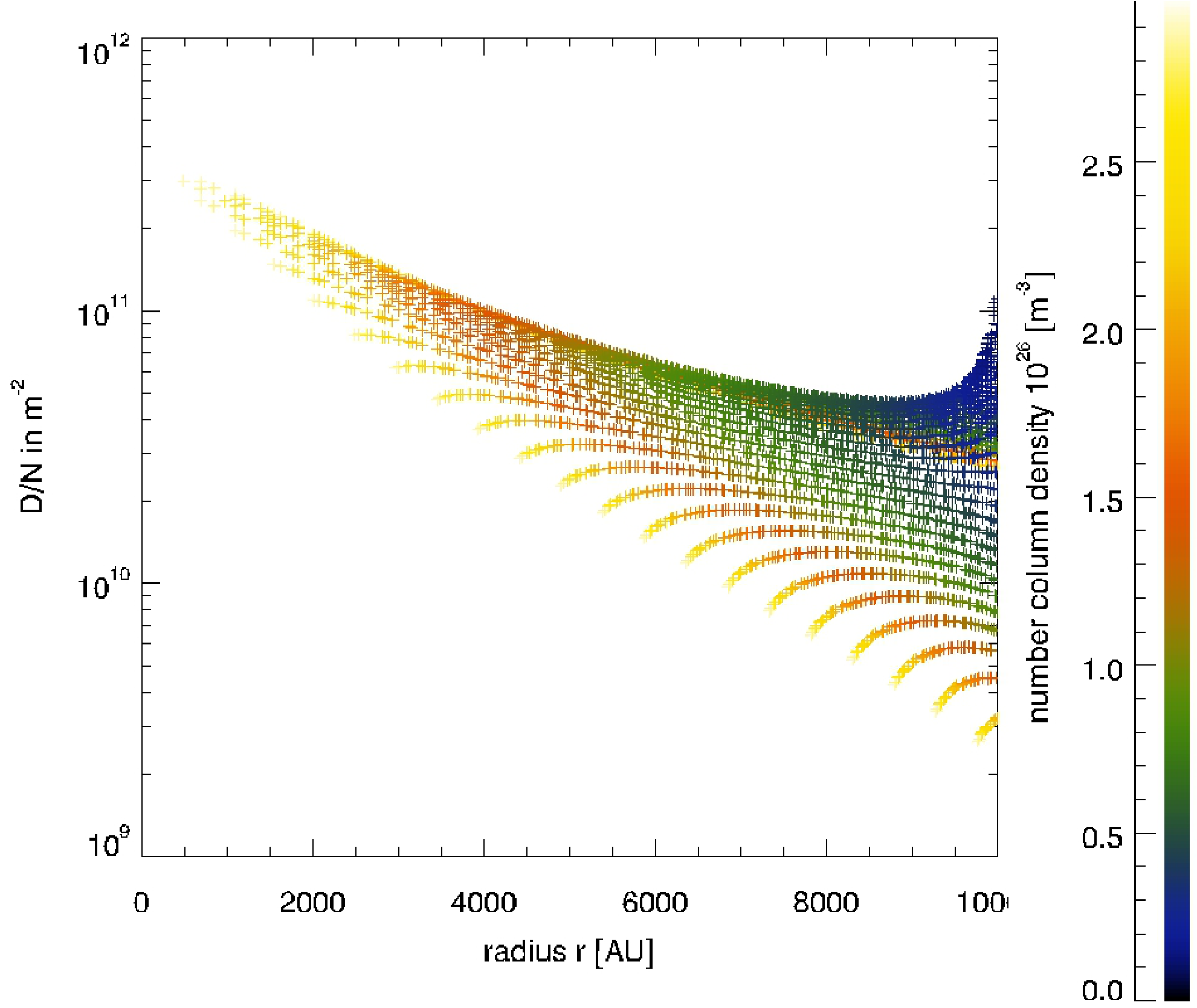}
\caption{
Directional transmission function normalized to the column density as a
function of the radius for grid cells in
a spherically symmetric standard molecular cloud core. The colors indicate 
the column density.
        }
\label{DN}
\end{figure}

For simple spherically symmetric cores with a radially decreasing density, the directional transmission function
has no strong impact on the detection of coreshine from cores with $\tau_{sca} <$ 0.7:
Fig.~\ref{DN} shows the ratio $D/N$ for grid cells
in a standard core with a radial number density profile of the form
\begin{equation}
n(r)=\frac{n_0}
      {1+\left(\frac{r}{r_0}\right)^\alpha}
\label{densi}
\end{equation}
inside $R_{out}=10,000$ au with the density normalization $n_0$ corresponding
to a core mass of 1 $M_\odot$, 
the radius of the flat inner region $r_0=1000$ au where 
the profiles changes from flat to power-law, and the power-law index
$\alpha=1.6$.
Since typical mean extinction cross sections for 1 $\mu$m-sized silicate grains
are of the order of 10$^{-16}$
m$^2$, estimating the contribution of each factor, and using the
upper limit derived from Fig.~\ref{DN}, we find that 
the second extinction correction term on the right hand side of Eq.~(\ref{FL2})
can be neglected compared to the first-order term for this type of core
at $\lambda=$3.6 $\mu$m.

For grains with $a>0.5$ $\mu$m, the correction term for the extinction 
can substantially contribute since
the grain cross section for extinction is dominated by the scattering
process.
Correspondingly, the 2nd-order balance region with zero coreshine
detection in Fig.~\ref{Flatline} (bottom)
encloses a smaller region than the 1st-order balance region for zero
coreshine in the middle panel of Fig.~\ref{Flatline}.

Finally, we note that the used balance limits do not include the sensitivity limit
of a particular instrument like that of IRAC band1 in the cryo or post-cryo phase.
To be detectable by that instrument the coreshine surface brightness must be larger than
the background surface brightness and the limit of the instrument. 
For a post-cryo measurement of the surface brightness of coreshine in LDN260, e.g.,
the brightness maximum is about a factor of 18 above the sensitivity limit, but
this additional detection barrier can increase the dark zone of coreshine detection.

\section{Radiative transfer calculations for scattered light from a model core}\label{RT}
To test the approximate scattering-extinction balance border, we have performed a grid of radiative transfer
calculations for a model core placed in various regions of the sky. 
We model the scattered light at 3.6 $\mu$m from a core with the density described by Eq.~(\ref{densi}),
a radius of 10,000 au and a mass of 1 $M_\odot$ using a 3D single-­‐scattering radiative transfer code. The core density
follows a radial power-law with index -1.5 and a flattening radius of 2500 au.
Derived from the
general code applied in \citet{2010A&A...511A...9S}, the code utilizes size distribution-integrated
opacities \citep[for details, see Sect.~2.4 in][]{2013ARA&A..51...63S}.
The scattering of the ISRF by an MRN grain size distribution is calculated
with varying upper size limits. 
The grain absorption and scattering efficiencies are computed following the Mie theory for spherical grains. The cores are illuminated by the ISRF only, without any local enhancement.
The code has also been used to generate fits to the near-­‐IR and 3.6 $\mu$m 
surface brightness
profiles of the LDN260 core in \citet{2013A&A...559A..60A}. For each ray followed in the code, the optical depth is calculated, and
we found more than 95\% of the rays stayed within the limits of the optical thin absorption and scattering
for the assumed core and grain properties.

\begin{figure}
\includegraphics[width=9cm,angle=0]{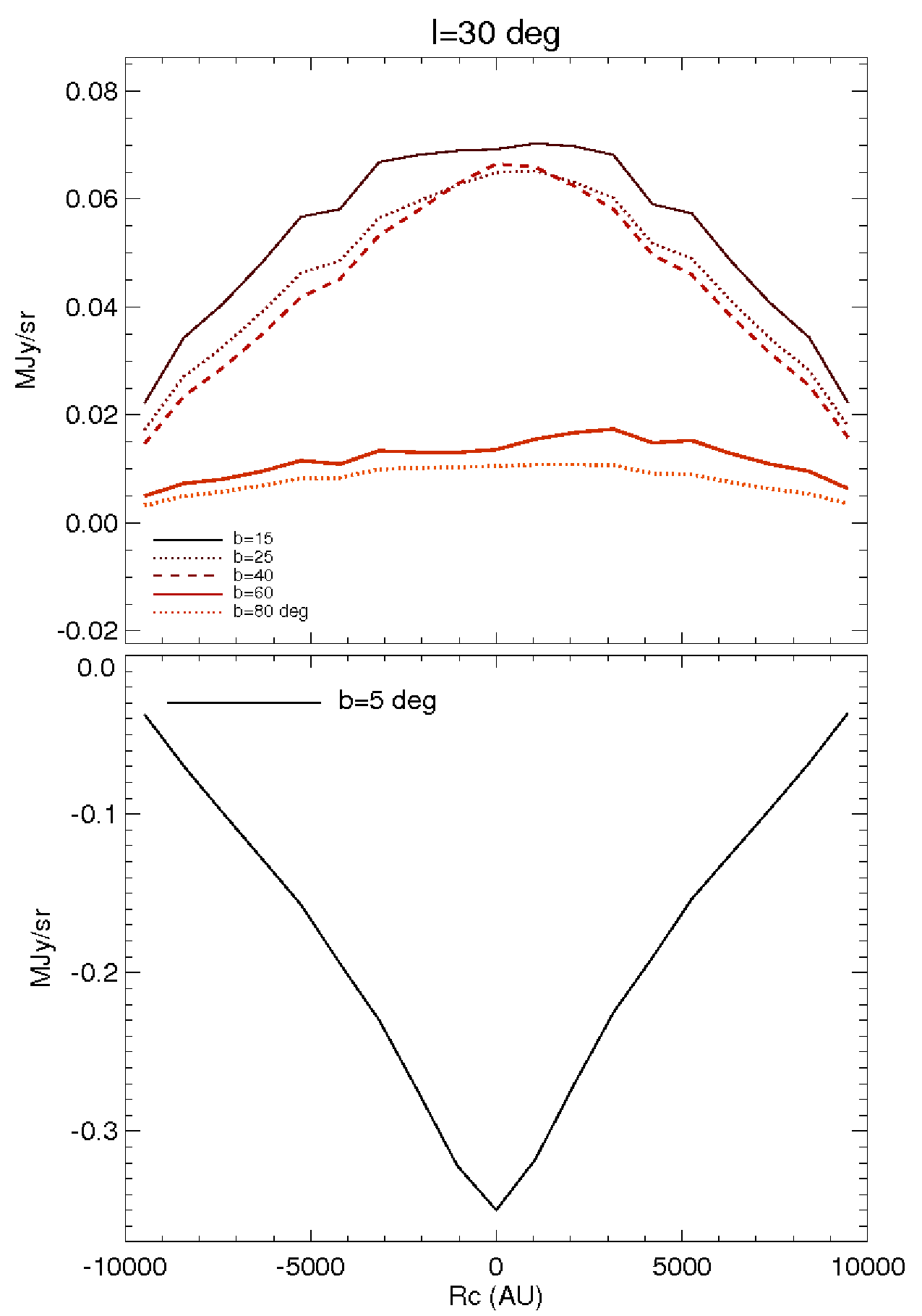}
\caption{
Surface brightness profiles at $\lambda=$3.6 $\mu$m of a spherically symmetric standard core
along a cut through the PoSky-projected core center position for a core
at various Galactic coordinates.
        }
\label{profiles}
\end{figure}

To illustrate the analytical detection criteria, we show in 
Fig.~\ref{profiles} 
the surface brightness cut through the center of a core located at 
l=30$^\circ$ and at various latitudes b from a LoS very close to Galactic
plane (5$^\circ$) to a high-latitude core at b=80$^\circ$. 
The background flux has been subtracted.
As expected absorption of the strong background dominates if the core is
located at 5$^\circ$ and the profile is seen in absorption. 
Cores located at higher latitudes are seen in excess compared to 
the background. 
The maximum excess is obtained for cores located between 20$^\circ$ and 40$^\circ$. 
Beyond 40$^\circ$, the scattering efficiency towards the observer is depressed by the phase function, which favors forward scattering.

\section{Investigating local effects for a set of example cores }\label{Cores}
The list of cores with potential coreshine candidates presented in
\citet{2010Sci...329.1622P} was based on a visual examination of the IRAC bands, searching
for an excess of surface brightness at 3.6 $\mu$m where extinction is found at 8 $\mu$m.
This was a sufficient approach to start the exploration of the global appearance of the
coreshine effect.
The strong local variation of the background, the possible appearance of PAH
emission in regions with a stronger local field like Rho Ophiuchi, or a weak
background at 8 $\mu$m causing a weak extinction contrast, however, makes a global assessment
of the appearance of coreshine in all observed cores difficult. 
The correct way therefore is to model the
cores individually including all local effects and to assemble a coreshine
detection list this way.

Therefore, instead of presenting an updated list of candidate cores which appear to show coreshine
based on a subjective "by-eye" criterion,
we discuss the impact of local effects for a sample of cores
in and outside of the Galactic plane. This is to prepare forth-coming detailed
core-by-core radiative transfer modeling based on the global detection criteria proposed in
this paper and discussing all local effects.

\begin{figure*}
\includegraphics[width=18.5cm,angle=0]{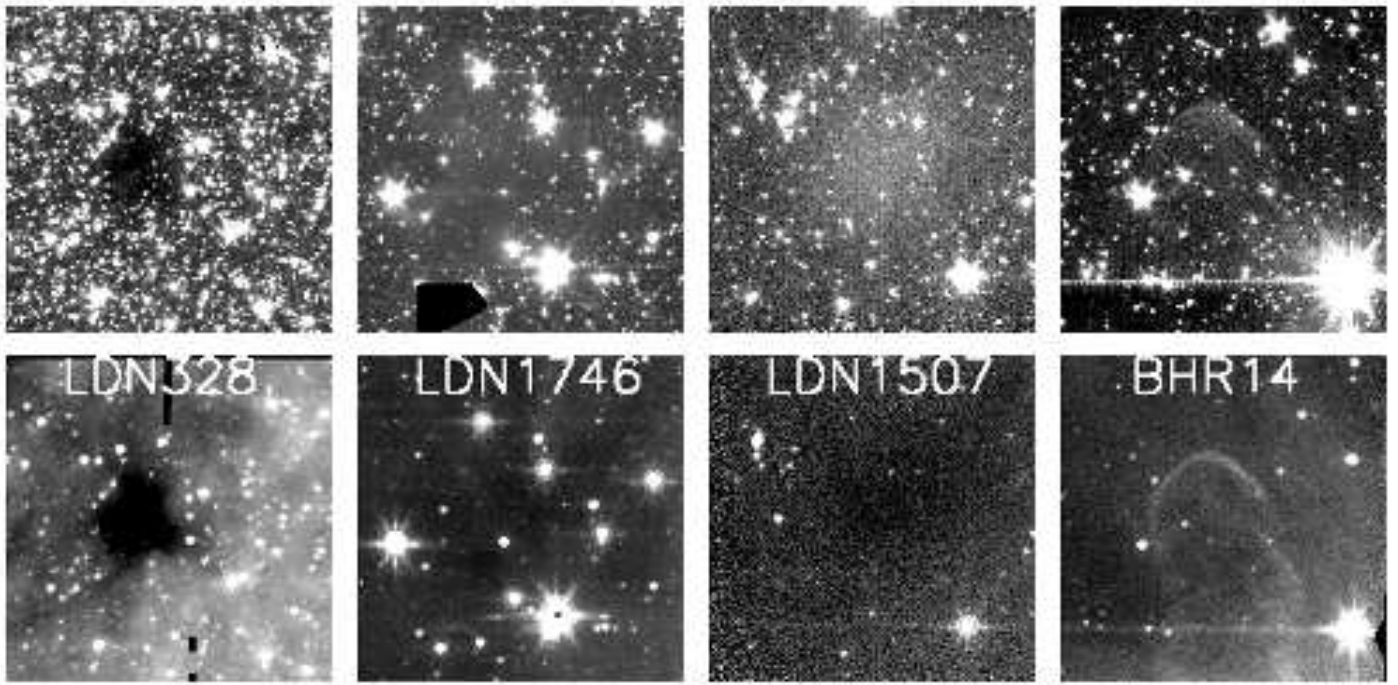}
\caption{
{\it Spitzer} 3.6 $\mu$m (top) and 8 $\mu$m (bottom) images of, 
from left to right, LDN328, LDN1746, LDN1507, and BHR14. 
North is up and east is to the left. 
Each image is 5\arcmin\ square. 
LDN1746 shows coreshine despite being located close to the Galactic plane and is predicted not to. 
However, the YSOs around LDN1746 are additional illumination sources. BHR14 shows clear surface
brightness excess at 3.6 $\mu$m but also PAH emission at 8 $\mu$m. 
} 
\label{obs_cores}
\end{figure*}

Fig.~\ref{obs_cores} shows {\em Spitzer} images for the four
example cores which we will discuss in the following. The upper images refer to 3.6 $\mu$m,
the lower 8 $\mu$m images are added as additional source of information.

LDN328 is an example of a core near the Milky Way midplane, close to the GC 
situated at (l,b)=(13.3$^\circ$,-0.83$^\circ$) where the background
is expected to be very high. 
Indeed, the COBE/DIRBE all-sky map provides a line of sight surface brightness 
of 2.4 MJy/sr after stars have been subtracted from each DIRBE pixel using the 
WISE point source catalog as discussed in
\citet{2013A&A...559A..60A}.
The coreshine from the model core used in Sect.~3 in the same region of the sky shows a strong
extinction profile and indeed, the left image of Fig.~\ref{obs_cores} shows the core in absorption.

LDN1746 is located above the GC (l,b)=(357.13$^\circ$, +07.17$^\circ$) and has a $b$ too low to 
be outside the main bulge of extinction.
The DIRBE/COBE star subtracted background is 0.4 MJy/sr which is stronger than the ISRF. 
However, LDN1746 (also called B59) is known to be the only core in the pipe nebula with 
young stellar objects (YSOs) close by. Future modeling of this complex source will reveal 
if they are close enough to provide a local ISRF as a source for the coreshine that is seen
in Fig.~\ref{obs_cores}.

LDN1507 in Taurus is conversely in a low background region below the anti-center.
Located at (l,b)=(171.51$^\circ$, -10.59$^\circ$) the star subtracted 
background is only 0.04 MJy/sr. 
Backscattering of most of the Milky Way radiation plus the contribution from the outer Milky Way in this
direction is therefore sufficient for coreshine to be 
brighter than the background, as observed.

BHR14 is situated in the Gum/Vela region (253.82,-10.9). Clear excess of surface brightness is seen at 3.6 $\mu$m. 
However, excess is also evident at 8 $\mu$m suggesting strong PAH emission which is also contaminating the 3.6 $\mu$m bandpass. This in general makes it difficult to identify coreshine compared to the PAH emission. 

\section{Concluding summary} \label{conclusions}
We find that the detection of scattered light at 3.6 $\mu$m from low-mass
molecular cloud cores is both influenced by local effects like the radiation
field in the vicinity of the core and global effects like the interplay
of core position with respect to the GC and 
the directional characteristics of the scattering.
For optically thin radiation and a constant grain size distribution throughout
the core, we derive a simple limit for detecting coreshine that holds for grains
with sizes smaller than 0.5 $\mu$m. 
The extinction of background radiation by the core prohibits detection
in large parts of the Galactic plane and especially near the LoS towards the GC.
For grain distributions extending to 1 $\mu$m, the forwardly-peaked
directional characteristics of the scattering favors the
detection of coreshine near the GC.
After combining the two effects, we identify regions above and below the
GC and anti-center to be those with enhanced detection.
Two further quantities enter the scattering-extinction balance formula: the local background and
the local ISRF. In this work we use the only all-sky map having an absolute flux
calibration which is the DIRBE map. 
Due to the coarse Healpix grid it provides only a mean value where strong deviations can
occur locally. Moreover, it contains the stellar contribution which is not
included in the core background modeling.
Another uncertainty is the local incident radiation field that is observed in
star formation regions and that is
likely produced by the warm dust near the forming stars. Its directional variation,
flux, and spectrum is unknown however, and very difficult to measure.
We expect that only individual core modeling may derive constraints on the
local field: for the core LDN260 modeled in 
\citep{2013A&A...559A..60A}
an enhancement of the incident
radiation field was necessary to reproduce the observed coreshine.
The derived scheme can be used to identify cores 
which deviate from the
expected flux and might be of interest
in terms of nearby radiation sources, PAH contamination, or grain properties.
Combining the global constraints
on coreshine detection, cores above or below the GC are best detectable.
The moderate amplification by backward scattering of GC radiation along
with the lower background supports detection of coreshine near the anti-center.

The opacities by \citet{1984ApJ...285...89D} used in this paper
have served as reference data for decades
as much as later work showed in how much modification in the chemical composition,
shape, size distribution will alter the values.
It remains true however that we have little observational information on the
exact properties of larger grains which leaves ample room
for changing the dust properties and distribution and arriving at the same effects.
Therefore, instead of investigating the impact of various dust model assumptions
on the derived global effects, we expect progress from modeling multi-wavelength
data for a large number of single cores
as presented in
\citet{2010A&A...511A...9S} and \citet{2013A&A...559A..60A} where dust modeling
is better constrained.
Future work also will include results from coagulation calculations
\citep{L1506C}.

\begin{acknowledgements}
JS, MA and WFT acknowledge support from the ANR (SEED ANR-11-CHEX-0007-01).
We thank L. Pagani, M. Juvela, R. Paladini, and C. Lefevre for helpful discussions, and the
referee for the constructive comments which helped to improve the manuscript.

\end{acknowledgements}

\bibliographystyle{aa}
\bibliography{Detectability}

\end{document}